\documentclass[prl,twocolumn]{revtex4-1}

\usepackage{amsmath}  
\usepackage{amsfonts} 
\usepackage{graphicx} 

\begin{document}


\title{Phase Cascade Bridge Rectifier Array in a 2-D lattice}

\author{M. Nazari}
\affiliation{Electrical and Computer Engineering Department, Boston University}

\author{A. M. Gole}
\affiliation{Electrical and Computer Engineering Department, University of Manitoba, Winnipeg}
\author{M. K. Hong}
\author{P. Mohanty}
\author{S. Erramilli}
\email{Corresponding authors: narayan@ucsc.edu and shyam@bu.edu}
\affiliation{Physics Department, Boston University,590 Commonwealth Avenue, Boston, MA 02215, USA}
\author{O. Narayan}
\email{Corresponding authors: narayan@ucsc.edu and shyam@bu.edu}
\affiliation{Physics Department, University of California Santa Cruz}


\date{\today}

\begin{abstract}
We report on a novel rectification phenomenon in a 2-D lattice network consisting of $N\times N$ sites with diode and AC source elements with controllable phases. A phase cascade configuration is described in which the current ripple in a load resistor goes to zero in the large $N$ limit, enhancing the rectification efficiency without requiring any external capacitor or inductor based filters. The integrated modular configuration is qualitatively different from conventional rectenna arrays in which the source, rectifier and filter systems are physically disjoint.  Exact analytical results derived using idealized diodes are compared to a realistic simulation of commercially available diodes. Our results on nonlinear networks of source-rectifier arrays are potentially of interest to a fast evolving field of distributed power networks.
\end{abstract}

\maketitle 

\section{Introduction} 


The need for DC  electrical power generation from AC sources is arguably the most important nonlinear problem in modern society.  Interest in distributed energy networks\cite{Blaabjerg2006}, alternative energy technologies \cite{Dresselhaus2001,Mateen2016} and the need for energy harvesting\cite{Alavikia2015}  and energy scavenging \cite{Qin2008} systems has spurred renewed interest in the problem.  A very large number of configurations have been investigated for power generation from AC source networks, and extend to wireless power generation \cite{Popovic2013,Alavikia2015}, rectenna arrays\cite{Olgun2011,Hagerty2004} and smart grids\cite{Farhangi2010}. Of particular interest is the generation of DC power, not only for its ubiquiitous use but also for efficiency in transmission over long distances with asynchronous power generators with a wide range of sources for energy harvesting.   A classic configuration consists of  a rectenna\cite{Brown1984} in which diode arrays are used to rectify the output of a receiving antenna, followed by filters to provided the desired DC power. Such nonlinear networks have become very interesting recently in models in which diode elements are integrated into plasmonic structures, in the rapidly emerging new field of nonlinear metamaterials\cite{Poutrina2010,Lapine2014}, among others. The ability to design and engineer local phase shifts within each element of a receiving array represents a new capability that has not been studied at all, and possible configurations remain underexplored. Lattice networks consisting of nonlinear elements like diodes have been extensively studied  in statistical physics, in directed networks and percolation problems\cite{redner:1982}, and random circuit element networks\cite{Karpov2003,Dearcangelis1985,Derrida1982}. The range of phenomena is potentially vast even for deterministic systems. Up to now, the inclusion of voltage sources within the lattice sites has not been 	considered. It may be expected that integration of sources with random or varying phases at each lattice site will lead to an even richer range of phenomena. Distributed generation of power from both conventional and alternative energy sources provides motivation to study networks in which voltage sources are also distributed throughout the lattice.

		In this article, we investigate rectification phenomena in a particular 2-D lattice network consisting of $N^2$ sites in which each lattice site consists of an AC source in a bridge-rectifier like configuration. Each lattice site shares one diode with each of its nearest neighbors, as shown in Figure~\ref{fig:wampum}. The configuration is inspired by rectenna arrays, but now with the addition of AC sources at each lattice site.  The system consists of $N^2$ AC sources, and $(N+1)^2$ diodes. It can be seen that when $N=1$, the configuration is identical to a classic Bridge rectifier combination. The current through the load resistor has many frequency components, with the zero frequency DC component being the most important for rectification.  Our results show that an exact analytical result valid for arbitrarily large $N$ can be derived even for such a strongly nonlinear system with idealized diodes.

\section{Models}

\begin{figure}[htb]
\begin{center}
\includegraphics[width=3in]{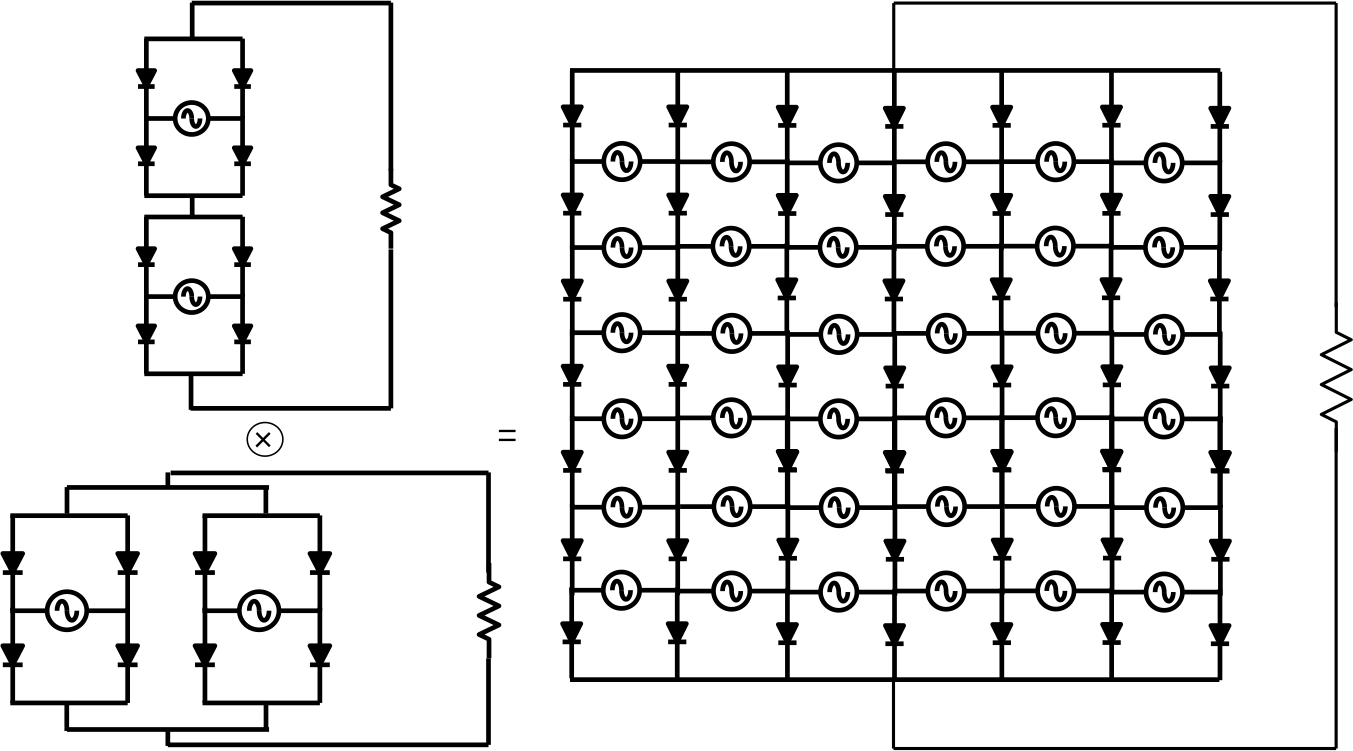}
\caption{{\it a:} Full-wave rectifiers arranged in series and parallel.  {\it b:} $N^2$ AC voltage sources and diodes arranged in a 2-D square lattice. }
\label{fig:wampum}
\end{center}
\end{figure}

We consider how to arrange a lattice of $M\equiv N^2$ AC voltage
sources, with suitably chosen relative phases and with
diodes for rectification, to produce an almost dc source. For definiteness, consider a 2-D lattice of elements arranged as shown on the right side of Figure~\ref{fig:wampum}. In the next section, we present the 
analysis for idealized diodes, which operate in just two states determined by the bias voltage across the diode: (i)  an ``off' state, at negative bias, where the current is zero; (ii) an ``on'' state,  at positive bias, where the current is an arbitrary positive value limited only by the load. In the following section, the results are compared to  the results of simulations using commercially available Silicon diodes with a forward bias set by the $\sim 0.7 V$ bandgap of Silicon, and current set by a non-zero forward resistance.  

For comparison, we extend the series  arrangement shown on the left side of Figure~\ref{fig:wampum} to the case where
$N^2$ voltage sources are arranged in series with a load resistor $R_L$. The total number of diodes required is $4M\equiv 4N^2$. Each diode must then be rated to carry up to the maximum total current flowing through the load resistor, which limits the utility of the series configuration.
The phases of the voltage
sources are uniformly distributed between zero and $2\pi.$ With the 
rectified sources in series, the total voltage at time $t$ is 
\begin{equation}
V(t) = \sum_{j=1}^{N^2} V_0 |\sin(\omega t + 2\pi j/N^2)|.
\end{equation}
Clearly, this is a periodic function of time, with time period $T = 2\pi/(\omega N^2).$ For the simple case when $N$ 
is an even number, we have
\begin{eqnarray}
V(\omega t = 2\pi k/N^2)  &=& 2 V_0 \cot(\pi/N^2) \nonumber\\
V(\omega t = \pi (2 k - 1)/N^2) &=& 2 V_0\csc(\pi/N^2)
\end{eqnarray}
are the minima and maxima of $V(t).$ 


The ratio of minimum to maximum voltage is $\cos(\pi/N^2),$ 
and the frequency of these ripples is $N^2 \omega.$ 
For large $N,$ the ripples have height $\pi V_0/N^2,$ and are superimposed
on a dc voltage of $2N^2 V_0/\pi.$
We note that the peak current carried through each diode is approximately $2 N^2 V_0/(\pi R_L)$, which it carries for half the cycle. For large $N$ this current can exceed the maximum rating for the diode in the series configuration.

The current load in each diode can be reduced by adopting a parallel configuration (see the left side of Figure~\ref{fig:wampum} ). For $M$ bridge rectifiers arranged in a parallel configuration, the sources have to be in phase. The fraction of the ripple in the output voltage can be seen to be just as large it would for a single rectifier, and additional filtering would be needed to get DC current through the load resistor. 

These elementary statements motivate us to investigate a hybrid system that combines series and parallel combinations  into the 2D lattice shown in Figure ~\ref{fig:wampum}, with each unit cell is mapped onto a conventional bridge rectifier.
Remarkably, we find that for a special choice of phases of the AC sources, the configuration provides the advantage of reduced ripple in the rectified DC voltage of a series combination, with the advantage of reduction in peak current that is the hallmark of the parallel combination.

\section{Analytical results}
\label{sec:analytical}
In this alternative arrangement shown on the right side of Figure~\ref{fig:wampum}, diodes are placed on the vertical sides of the square, and 
voltage sources on the horizontal sides. Conducting strips at the top
and the bottom are linked to the external load. The voltage across 
the source in the $i$'th row and $j$'th column is 
\begin{equation}
v_{ij}(t) = (-1)^i V_0 \sin\left[\omega t + 
2\pi \left(\frac{j}{N} +\frac{i}{N^2}\right)\right]
\label{brick_sources}
\end{equation}
where $0 \leq i \leq N - 1$ and $1 \leq j \leq N.$ The sign of the voltage
across a source is fixed by defining it as the difference between the
voltage at the right end and at the left end of the source.  Then one can
verify that the voltage at the $j$'th junction between voltage sources
in the $i$'th row is
\begin{equation}
V_{ij}(t) = \frac{(-1)^{i-1} V_0}{2\sin (\pi/N)} 
\cos\left[\omega t + 2\pi \left( \frac{j + 
\frac{1}{2}}{N} + \frac{i}{N^2}\right)\right] + c_i(t)
\label{brick_nodes}
\end{equation}
where $0 \leq j \leq N$ and $c_i(t)$ is a $j$-independent function of 
time in the $i$'th row
that has to be determined. (Eq.(\ref{brick_nodes}) can be verified
by taking the difference in the voltages at successive nodes and
comparing to Eq.(\ref{brick_sources}).) The voltages $V_{ij}$ are
then the voltages at the ends of the various diodes. The functions
$c_i(t)$ can be determined by the conditions that, for idealized diodes, 
i) the voltage across each diode must be negative or zero ii) at least
one diode per row must have a non-negative voltage across it so that 
current can flow from the row of voltage sources above it to the row
below. This ensures that 
\begin{equation}
\min_j[V_{i+1,j}(t) - V_{ij}(t)] = 0
\label{minmax}
\end{equation}
indepdent of $t,$ 
for $N - 1 > i \geq 0.$ This fixes $c_{i+1}(t) - c_i(t).$ 
Similar reasoning can determine the voltages at the conducting
strips at the top and bottom of the lattice.

To understand the results obtained, we first consider a
 continuum approximation, where the number of nodes in each row
is very large. 
From Eq.(\ref{brick_nodes}), the voltage across
any row is a sinusoidal curve with amplitude $N V_0/(2\pi).$
The sinusoidal curve for each row is upside down compared to the adjacent rows. 
(The curve in each row is also shifted
horizontally by a phase of $2\pi/N^2$ with respect to the preceding row, but 
this shift is neglegible in the  $N\rightarrow\infty$ limit.) From Eq.(\ref{minmax}), we 
immediately obtain 
\begin{equation}
c_{i+1}(t) - c_i(t) \rightarrow  - \min_j \frac{2NV_0 (-1)^{i-1}}{2\pi}\cos\left(\frac{2\pi j}{N}  +\phi_i(t)\right) 
\end{equation}
where $\phi_i(t) = \omega t + \pi/N + 2\pi i/N^2.$ Thus $c_{i+1}(t) - c_i(t) = N V_0/\pi.$
By the same reasoning, the voltages of the two conducting strips are $c_0(t) - N V_0/(2\pi)$ and $c_{N-1}(t) + N V_0/(2\pi).$ 
Therefore the voltage difference 
between the two conducting strips is $N^2 V_0/\pi,$
and a DC current flows across the load resistor. 

With an understanding of the behavior of the ideal phase cascade shown in the $N\rightarrow \infty$ limit, we now consider the case of a finite lattice, with $N$ an even number for simplicity.  The displacement in the sinusoidal
curves for successive rows, i.e. $c_{i+1}(t) - c_i(t),$ is obtained by the 
condition
\begin{widetext}
\begin{equation}
\min_j \frac{(-1)^i V_0}{2\sin(\pi/N)} \left\{
\cos\left[\omega t + 2\pi\left(\frac{j + \frac{1}{2}}{N} + 
\frac{i + 1}{N^2}\right)\right]
+\cos\left[\omega t + 2\pi\left(\frac{j + \frac{1}{2}}{N} + 
\frac{i}{N^2}\right)\right] \right\} + c_{i+1}(t) - c_i(t) = 0.
\label{cici1}
\end{equation}
\end{widetext}
Because $N$ is even, replacing $j\rightarrow j + N/2$ changes the sign
of both cosine terms. Since the minimum over all $j$ is taken, the 
factor of $(-1)^i$ can be dropped.
We can also replace the minimum with the maximum, with a change of sign, with $\Delta c_i\equiv c_{i+1}(t) - c_i(t) $:

\begin{equation}
\Delta c_i = 
\max_j\frac{V_0\cos(\pi/N^2)}{\sin(\pi/N)} 
\cos\left[\omega t + 2\pi\left(\frac{j + \frac{1}{2}}{N} + 
\frac{i + {1\over 2}}{N^2}\right)\right]
\label{cici1b}
\end{equation}

The voltage of the conducting strip at the top  of Figure~\ref{fig:wampum} is
\begin{equation}
V_{top}(t) = c_0 -  \max_j\frac{V_0}{2\sin(\pi/N)}
\cos\left[\omega t + 2\pi\left(\frac{j +
\frac{1}{2}}{N} \right)\right]
\end{equation}
where we have 
used the fact that $j\rightarrow j + N/2$ reverses the function. The voltage of
the conducting strip at the bottom of Figure~\ref{fig:wampum} is  similarly
\begin{equation}
V_{bottom} = c_{N-1} + \max_j \frac{V_0}{2\sin(\pi/N)}
\cos\left[\omega t + 2\pi\left(\frac{j + \frac{1}{2}}{N} -
\frac{1}{N^2}\right)\right]
\end{equation}
where we have replaced $j$ with $j-1$ to absorb a phase of $2\pi/N$ from $2\pi i/N^2.$ 
The voltage difference between the two conducting strips can be obtained by subtracting these two equations. 
It is convenient to write this as 
\begin{equation}
\Delta V(t) = c_N(t) - c_0(t) + \delta V(t)
\label{DeltaVt}
\end{equation}
by formally extending Eq.(\ref{cici1}) to $i=N-1$ with $j\rightarrow j - 1,$ where 
\begin{eqnarray}
\delta V(t) &=&  \frac{V_0}{2\sin(\pi/N)} 
\Bigg(
\max_j\cos\left[\omega t + 2\pi\left(\frac{j + \frac{1}{2}}{N} -
\frac{1}{N^2}\right)\right] \nonumber\\
&+& \max_j
\cos\left[\omega t + 2\pi\left(\frac{j +
\frac{1}{2}}{N}
\right)\right]
\nonumber\\
&-& \max_j \bigg\{
\cos\left[\omega t + 2\pi\left(
\frac{j + 1/2}{N} 
\right)\right]\nonumber\\
&+&\cos\left[\omega t + 2\pi\left(
\frac{j + \frac{1}{2}}{N} - \frac{1}{N^2}
\right)\right] 
\bigg\}
\Bigg).
\label{deltavt}
\end{eqnarray}

For the first part of Eq.(\ref{DeltaVt}), using Eq.(\ref{cici1b}), 
\begin{widetext}
\begin{equation}
c_N(t - \tau) - c_0(t-\tau) = 
\frac{V_0\cos(\pi/N^2)}{\sin(\pi/N)} \sum_{i=-N/2}^{N/2-1}\max_j 
\cos\left[\omega t + 2 \pi\left(\frac{j}{N} 
+ \frac{i}{N^2}\right)\right]
\label{summ}
\end{equation}
\end{widetext}
where we define $\omega\tau = \pi/N + \pi/N^2,$ and we have used the periodicity of the 
summand with respect to $i\rightarrow i + N$ to change the limits of the sum. 
This is a periodic function of time, with a period  of $2\pi/(\omega N^2).$
If $0 < \omega t < 2\pi/N^2,$ which covers one time period, every 
term in the sum on the right hand side of Eq.(\ref{summ}) is maximized at $j=0.$ Thus the minima and maxima
of $c_N(t-\tau) - c_0(t-\tau)$ occur at 
\begin{equation}
c_N(-\tau) - c_0(-\tau) = 
\frac{V_0\cos^2(\pi/N^2)}{\sin(\pi/N^2)}
\end{equation}
and 
\begin{equation}
c_N\left(\frac{\pi}{\omega N^2} - \tau\right) - c_0 \left(\frac{\pi}{\omega N^2} -\tau\right) = 
V_0\cot(\pi/N^2).
\end{equation}
The ratio of the 
two is $\cos(\pi/N^2).$ Thus for large $N,$ $c_N (t) -
c_0(t)$ tends a dc voltage of magnitude $N^2 V_0/\pi,$ with ripples
whose frequency is $N^2\omega$ and height is $\pi V_0/(2 N^2).$

We now turn to $\delta V(t).$ From Eq.(\ref{deltavt}), 
$\delta V(t)$ is periodic with a period
$2\pi/(N\omega).$ Furthermore, $\delta V(t) = 0$ unless the two cosine
functions have their maxima at different values of $j.$ The first cosine
has its maximum at $j=0$ for $ -2\pi/N < \omega t - 2\pi/N^2 < 0 ,$
while the second cosine has its maximum at $j=0$ for 
$-2\pi/N < \omega t < 0.$ Thus $\delta V(t)$ is non-zero within 
the narrow time interval $0 < \omega t < 2\pi/N^2,$ or more generally,
$0 < \omega t + 2 m \pi/N < 2 \pi/N^2$ for integer $m.$ 
The maxima 
of $\delta V(t)$ occur when $\omega t + 2 m \pi/N = \pi/N^2.$ It is easy 
to evaluate Eq.(\ref{deltavt}) at $\omega t = \pi/N^2$ and verify that 
\begin{equation}
\delta V_{max} 
= V_0 \sin(\pi/N^2).
\end{equation}
Combining with the result of the previous paragraph, $\Delta V(t)$ has
a ripple of height $\pi V_0/(N^2)$ and frequency $N\omega$ and a ripple
of height $\pi V_0/(2 N^2)$ and frequency $N^2\omega,$ superimposed on
the dc voltage $N^2 V_0/\pi.$

We now compare the square lattice with $N^2$ voltage sources to $N^2$ full-wave
rectifiers in series. For a distributed voltage source, it is reasonable to 
choose the amplitude of the individual sources to scale as $\sim 1/N,$ i.e. 
$V_0 = V_c/N.$ Then, the DC voltage for the square lattice is $N V_c/\pi,$
compared to $2 N V_c/\pi$ for the rectifiers in series. On the other hand, 
for large $N$ there is only one diode per source instead of four. As mentioned 
earlier, for a load $R_L$ the maximum current flowing through each diode is 
$2 N V_c/(\pi R_L)$ for the series configuration. Similarly, for the 
square lattice with idealized diodes, the maximum current is $N V_c/(\pi R_L)$ 
because at any instant the current flows 
from top to bottom along a unique set of diodes. However, with realistic diodes,
the current is distributed over numerous parallel `channels', reducing the maximum
current. (Without numerical simulations, it is not obvious whether the number of 
channels $n$ is $O(1)$ or $O(N).$) Thus for the same current flowing through
the load resistor, it is easier to exceed the maximum rating of the diodes for 
rectifiers in series. The maximum voltage across a diode when it is reverse 
biased is $V_0 = V_c/N$ for the rectifiers in series. For the lattice, the 
maximum reverse biased voltage can be approximately obtained from the 
continuum approximation, as $2 N V_0/\pi = 2 V_c/\pi.$ Both of these are well behaved
for large $N.$ Finally, the power dissipated in the diodes is non-zero if they 
are not ideal. Since 
\begin{eqnarray}
I &=& I_s (\exp[V/V_s] - 1) \nonumber\\
  &\approx& I_s\exp[V/V_s] \qquad V > 0\nonumber\\
  &\approx& -I_s \qquad V < 0,
\end{eqnarray}
the instantaneous power dissipated is
\begin{equation}
\sim \sum_\alpha V_s I_\alpha \ln (I_\alpha/I_s) + \sum_\beta V_\beta I_s
\end{equation}
where the sums run over the forward and reverse biased diodes
respectively.  For the lattice, there are $O(N^2)$ reverse biased diodes 
with $O(1)$ voltages across them, and the power dissipated in them is 
$O(N^2).$ For the forward biased diodes, if the current
flows along $n$ channels, there are $O(nN)$ forward biased diodes each with a 
current $O(N/n)$ flowing through it. The power dissipated in the forward biased
diodes is thus $O(N^2\ln N).$ By comparison, for the rectifiers in series, 
there is an $O(N)$ current flowing in $O(N^2)$ forward biased diodes, so that
the power dissipated in them is 
$O(N^3 \ln N).$

Our results show  prefect rectification can be achieved using a nonlinear network consisting of a lattice of AC sources. This work provides an interesting example of an exact result that can be established for an infinite nonlinear network, consisting of ideal bridge rectifiers that are of interest for distributed energy generation.  In the next section, simulations with realistic diodes confirm the exact results derived here. Importantly, the phase cascade sequence described is independent of frequency, and perfect rectification can be achieved for a wide range of frequencies, which makes it attractive for energy harvesting and distributed energy networks. 

\section{Numerical simulations}
\label{sec:numerical}

\begin{figure}[htb]
\begin{center}
\includegraphics[width=3in]{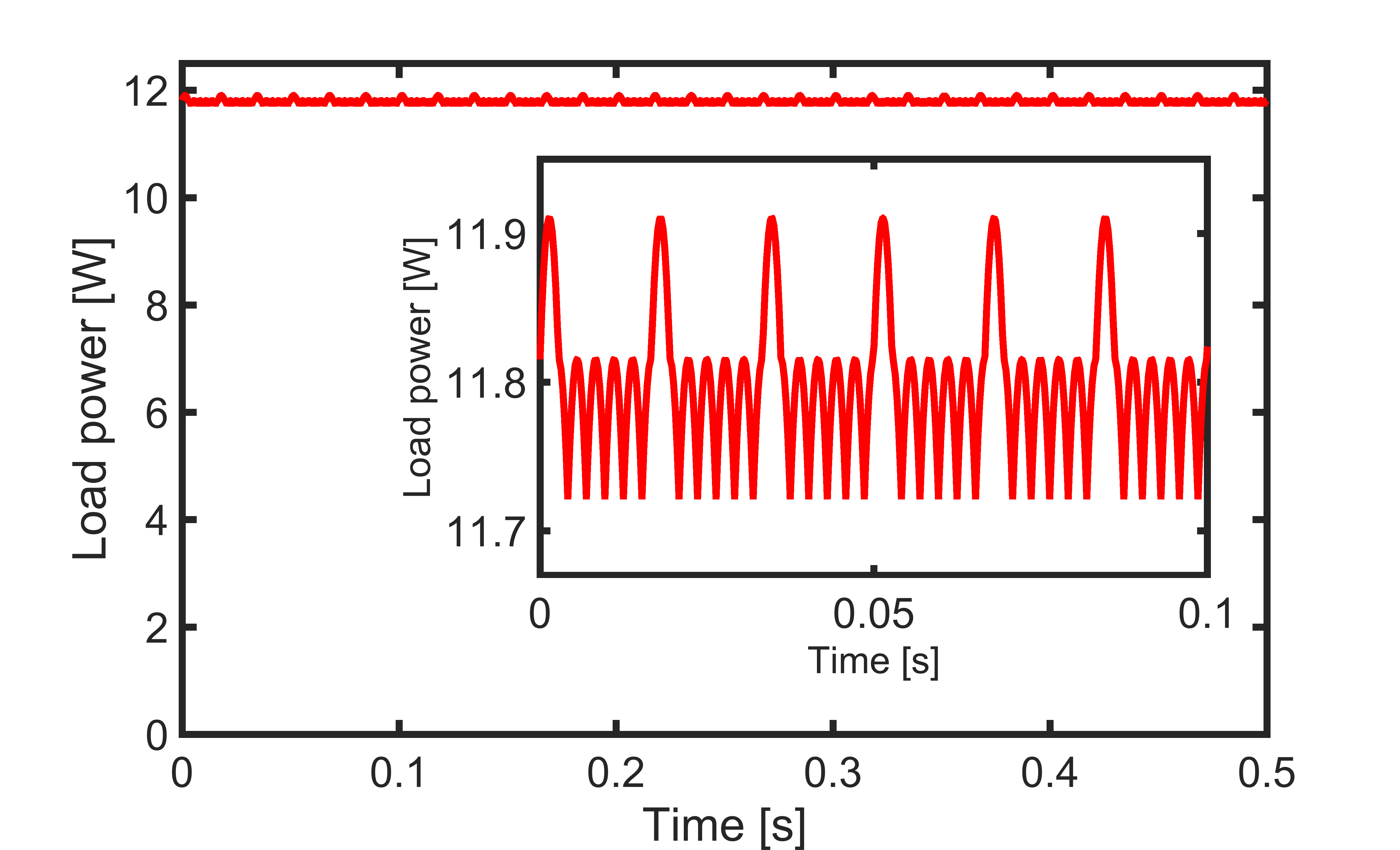}
\caption{Load Resistor power as a function of time for a $6\times 6$ array of 1N4004 diodes, with 5V AC sources operating at 50 Hz. The load resistor value has been set at 1 k$\Omega$. Inset shows an the same time series with an expanded Load Power scale.}
\label{timeseries}
\end{center}
\end{figure}

Simulations were performed in MATLAB Simulink, using Silicon diodes, with model parameters selected for a General purpose Rectifier Fairchild 1N4004). The amplitude of the AC sources was set at 10 V, and the frequency was set to 10Hz, driving a load a 1 k$\Omega$ load resistor. Current probes were used to measure the load current, and the current and voltage difference across selected diodes and AC power sources. The load current was used to derive the load power. Shown in Figure~\ref{timeseries} are the results of the simulation of the power for the 2-D square lattice with the Phase Cascade Array. \begin{figure}[htb]
\begin{center}
\includegraphics[width=3in]{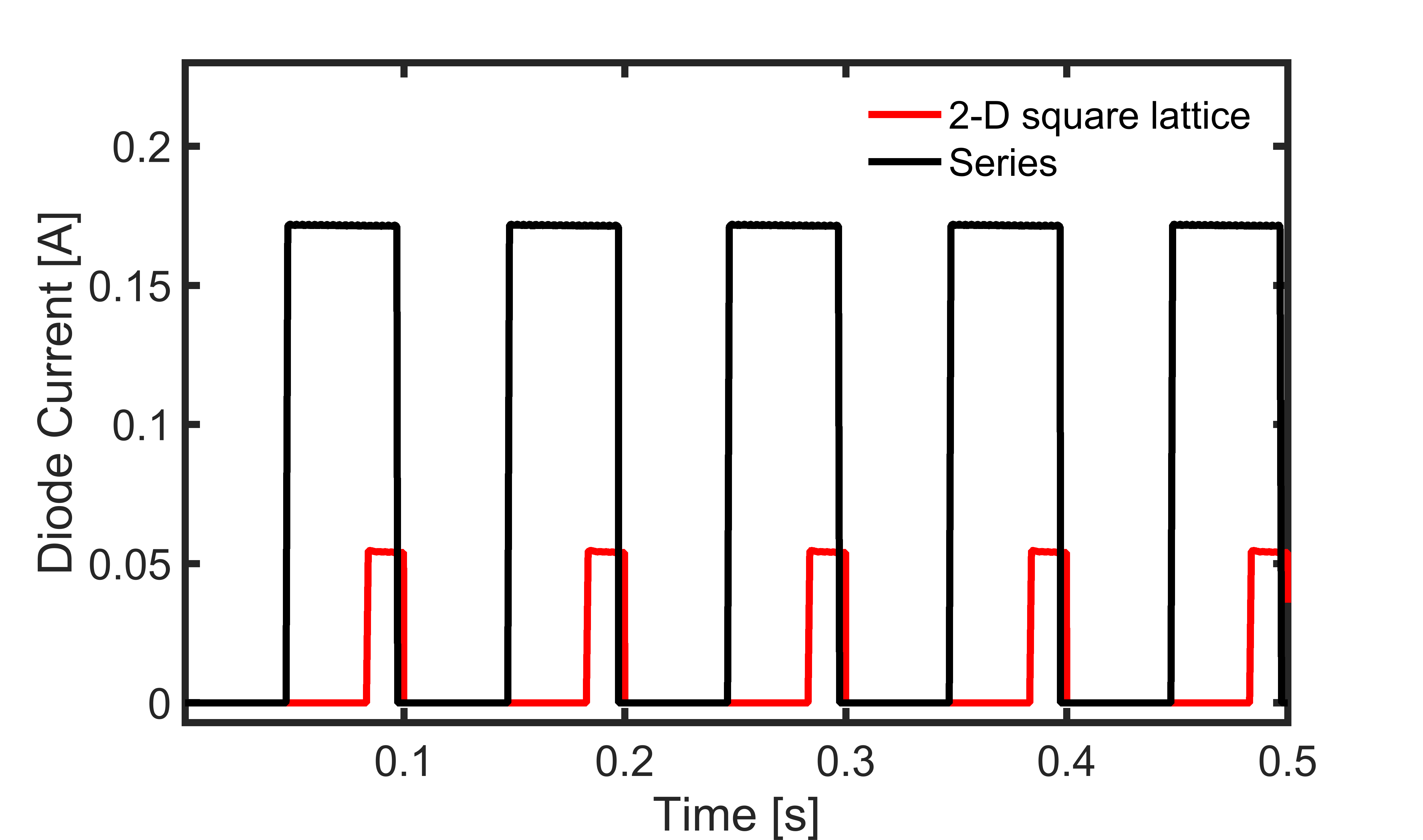}
\caption{(a) Current through a single selected diode in the series configuration compared to the Phase Cascade Array, with $N^2 = 6\times 6$.}
\label{dutycycle}
\end{center}
\end{figure}

\begin{figure}[htb]
\begin{center}
\includegraphics[width=3in]{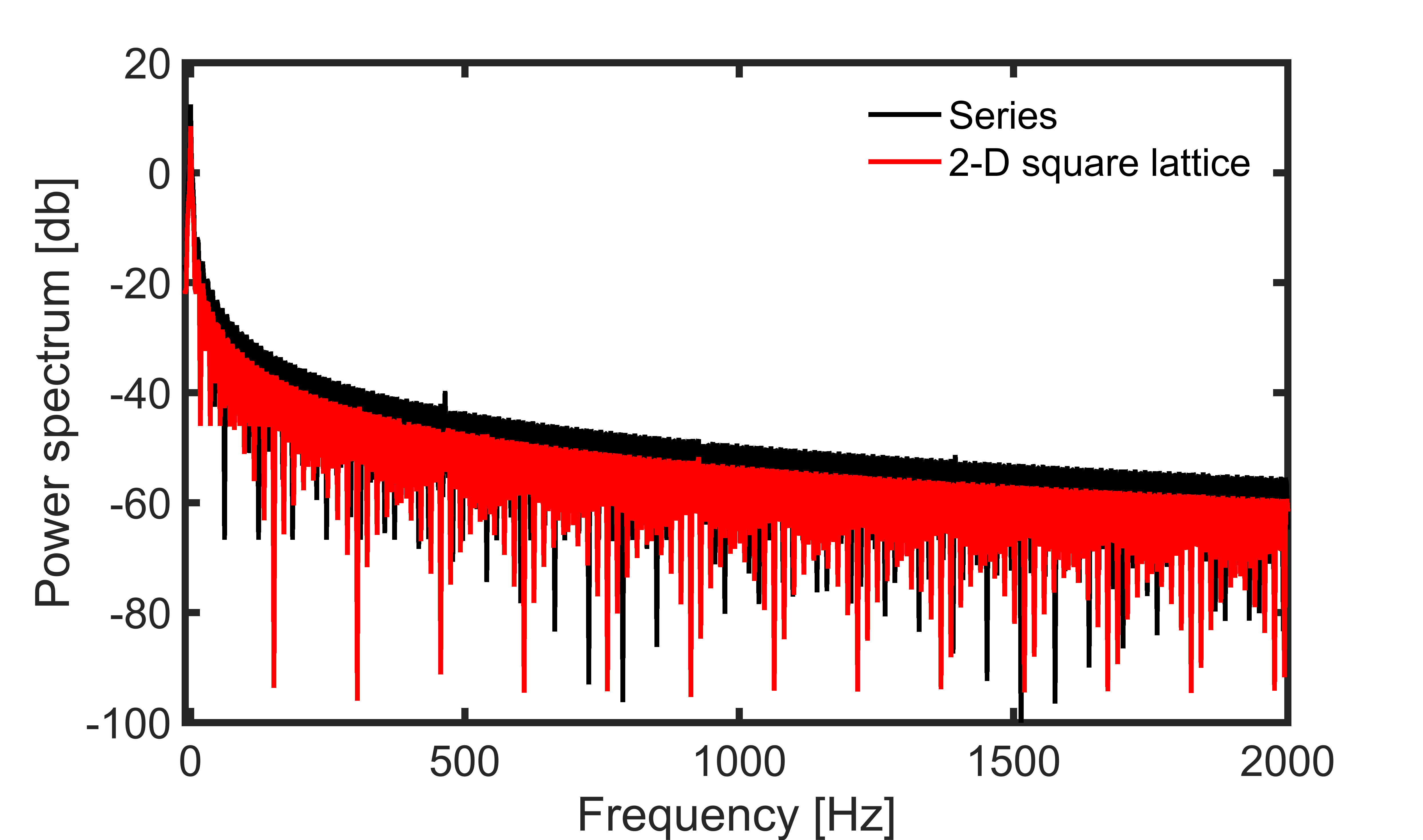}
\caption{ Fourier transform power spectrum. Red line corresponds to the power spectrum of the $6^2$ phase cascade rectifier array. 
The black curve is for the Series configuration. }
\label{fourier1}
\end{center}
\end{figure}

The simulations were used to calculate the representative duty cycle of the Phase Cascade Array, compared to the series configuration in Figure~\ref{dutycycle}. It is evident that in the Phase Cascade Array, the peak current is lower, and the duty cycle also lower than in the Series configuration, reducing both thermal load and easing the limits on the forward current. Fast Fourier Transforms of the time series were used to calculate the power spectrum in the load resistor. Figure~\ref{fourier1} shows a comparison of the power spectrum for the Phase Cascade Array compared to the series configuration.

\section{Acknowledgements}

\begin{acknowledgments}

M. Nazari acknowledges support from a Graduate Fellowship in the ECE department at Boston University. We thank C. Maedler, R. Averitt, and members of the Photonics Center staff for assistance.

\end{acknowledgments}

\end{document}